\documentclass[12pt]{iopart}

\usepackage{graphicx}
\usepackage{harvard}
\usepackage{braket}
\usepackage{xspace}
\usepackage{appendix}
\usepackage[export]{adjustbox}
\def\ppbar{\mbox{(anti-)}proton\xspace}

\def\ppbarss{\mbox{(anti-)}proton's\xspace}

\begin{document}

\title[Fast adiabatic transport]{Fast adiabatic transport of single laser-cooled $^9$Be$^+$ ions in a cryogenic Penning trap stack}

\author{T Meiners$^1$, J A Coenders$^1$, J Mielke$^1$, M Niemann$^1$, J M Cornejo$^1$, S Ulmer$^2$ and C Ospelkaus$^{1,3}$}

\address{$^1$ Institut für Quantenoptik, Gottfried Wilhelm Leibniz Universität, Hannover, Germany}
\address{$^2$ RIKEN, Ulmer Fundamental Symmetries Laboratory, Wako, Saitama, Japan}
\address{$^3$ Physikalisch-Technische Bundesanstalt, Braunschweig, Germany}
\ead{t.meiners@iqo.uni-hannover.de}
\vspace{10pt}
\begin{indented}
\item[]December 2022
\end{indented}

\begin{abstract}
High precision mass and $g$-factor measurements in Penning traps have enabled groundbreaking tests of fundamental physics. The most advanced setups use multi-trap methods, which employ transport of particles between specialized trap zones. Present developments focused on the implementation of sympathetic laser cooling will enable significantly shorter duty cycles and better accuracies in many of these scenarios. To take full advantage of these increased capabilities, we implement fast adiabatic transport concepts developed in the context of trapped-ion quantum information processing in a cryogenic Penning trap system. We show adiabatic transport of a single $^9\mathrm{Be}^+$ ion initially cooled to 2\,mK over a 2.2\,cm distance within\,15 ms and with less than 10\,mK energy gain at a peak velocity of 3\,m/s. These results represent an important step towards the implementation of quantum logic spectroscopy in the \ppbar system. Applying these developments to other multi-trap systems has the potential to considerably increase the data-sampling rate in these experiments.
\end{abstract}

\section{Introduction: file preparation and submission}

A stringent test of CPT (charge-parity-time) invariance and a sensitive probe to constrain physics beyond the Standard Model  is the comparison of the $g$-factor of the proton and the antiproton. Until today, the proton $g$-factor could be determined with a fractional precision of 0.3 ppb \cite{Schneider_2017} and the antiproton $g$-factor with a fractional accuracy of 1.5 ppb \cite{smorra_350-fold_2018}.  In these experiments the \ppbarss spin state was determined using the continuous Stern-Gerlach effect \cite{Dehmelt_1973}, \cite{smorra_base_2015}. An alternative approach to the continuous Stern-Gerlach effect is the determination of the \ppbarss spin state by applying quantum logic spectroscopy (QLS) with an atomic ion following a proposal by Heinzen and Wineland \cite{heinzen_quantum-limited_1990}. A key requirement for its realization are ions cooled to their motional ground state. This was demonstrated many times in RF traps, first by \cite{Diedrich_1989}, and later demonstrated in a Penning trap for the first time \cite{Goodwin_2016}. QLS was experimentally demonstrated in RF traps, first in \cite{Schmidt_2005}.

Because of conflicting requirements concerning the trap geometry in different steps of the spectroscopy algorithm, advanced multi-Penning trap stacks consisting of specialized zones interconnected through particle transport appear to be the most viable option to apply the QLS concept to the \ppbar system~\cite{Cornejo_2021}. Adiabatic transport of particles is a key ingredient for this approach. In the context of ion-trap quantum computing in Paul traps, transport of qubit ions is a key requirement to implement the ``quantum CCD architecture''~ \cite{wineland_experimental_1998,kielpinski_architecture_2002}. Near ground state transport in a linear geometry was first demonstrated by Rowe et al.~\cite{rowe_transport_2002} and later also through junctions~\cite{blakestad_near-ground-state_2011}. In radio-frequency traps much effort was invested in the past decade to implement, characterise and optimise fast ion transport with low heating rates. For reference see e.g. \cite{bowler_2013,Walther_2012,todaro_improved_2020,furst_controlling_2014,clark_characterization_2023}. Also in Penning traps, transport of ions is a key ingredient in many multi-trap methods. For example, the most precise comparisons of the fundamental properties of protons and antiprotons \cite{borchert202216,smorra_350-fold_2018}, mass measurements on highly charged ions with impact on neutrino physics, axion research, and potential future clock state characterization \cite{schussler2020detection}, as well as the most precise measurements of the masses of the electron \cite{sturm2014high} and the proton \cite{heisse2017high} rely on the execution of multi-trap measurement protocols.  Most of this work has primarily dealt with transport timescales of several seconds per protocol. While in some experiments heating rates as low as a few radial quanta per transport were achieved, in some cases the particle shutting consumes up to 10$\,\%$ of the measurement time budget and accumulates over the long data taking runs typically applied in these experiments to macroscopic time scales \cite{smorra_350-fold_2018}. 

In this article, we apply transport methods commonly used for adiabatic transport in RF traps to a single $^9$Be$^+$ ion in a Penning trap system, demonstrating for the first time in a Penning trap the transport of a single ion in the mK and ms regime. In addition to being a first step towards the single ion transport in the motional ground state required for the implementation of QLS, the work presented here has the potential to be applied to current high precision multi-trap Penning trap experiments by increasing the measurement sampling rate and providing results with lower statistical uncertainty. The outline of this paper is as follows. Section \ref{sec:theory} describes the method used for determining voltage waveforms for adiabatic transport which is a simplification of the method described in \cite{blakestad_near-ground-state_2011}. Section \ref{sec:setup} gives a brief overview over the experimental setup.  Section \ref{Sect:Experimental verification} describes the experimental implementation and demonstrates particle transport over a 2.2\,cm distance within\,15 ms and with less than 10\,mK energy gain at a peak velocity of 3\,m/s. 

\section{Adiabatic transport in Penning traps}
\label{sec:theory}

In a Penning trap a charged particle is confined by the superposition of an electrostatic quadrupole field and a homogeneous magnetic field \cite{DEHMELT196853}. The direction of the magnetic field is referred to as the axial direction. In this direction the particle oscillates at the axial frequency $\nu_z=\frac{1}{2 \pi} \sqrt{ 2 \mathrm{C}_2 V \frac{q}{m}}$ where $q/m$ is the charge-to-mass ratio of the confined particle, $V$ the voltage difference between the ring electrode and the endcaps, and $\mathrm{C}_2$ a coefficient describing the trap geometry \cite{gabrielse_open_1989}. The particle movement in the radial direction is described by the modified cyclotron frequency and magnetron frequencies, respectively, $\nu_\pm = \frac{1}{2\pi}(\frac{\frac{q}{m}B}{2}\pm \sqrt{\frac{(\frac{q}{m}B)^2}{4}-\frac{\nu_z^2}{2}})$ where $B$ is the magnetic field strength.

To adiabatically transport an ion through a cylindrical Penning trap along the axial direction, the axial trap frequency is kept fixed while the ion is moved along the axial direction. To accomplish this, time varying electrode voltages, which will be referred to as waveforms in the following, must be determined. To calculate the waveforms, it was assumed that the trap is rotationally symmetric, that the ion is located on the trap axis, and that its spatial spread is negligible. Therefore, the radial motions are not taken into account. These approximations reduce the problem to one dimension.

The waveforms are constructed from voltage vectors. The voltage vector $\vec{u}\vert_{z_0}$ contains the voltages that need to be applied to the individual electodes to trap the ion at positon $z_0$ (on the trap axis) i.e. form a harmonic potential at position $z_0$. The time dependence of the voltage waveforms is introduced by calculating voltage vectors for a set of ion positions along $z$ and then joining the resulting voltages for the individual electrodes choosing a parametrisation $z(t)$. The voltage vector for a particular ion position $z_0$ is determined by taking advantage of the superposition principle: One determines the potential produced at position $z_0$ by each individual electrode when 1~V is applied to this and 0~V to all the other electrodes. Based on these potentials and their derivatives with respect to $z$ a voltage vector $\vec{u}$ can be found that fulfills 
\begin{equation}
	A \cdot \vec{u} = \vec{b}
	\label{eq:Aub}
\end{equation}
where $A$ contains the electrode potentials and their derivatives and $\vec{b}$ contains the constraints for the electric potential and its derivatives at the ion position. $\vec{b}$ is invariant for all individual ion positions. This method is an adaption of the method presented in \cite{blakestad_near-ground-state_2011} for radio-frequency traps to the 1-dimensional and pseudopotential-free case of the Penning trap. 

\begin{figure}[ht]%

	\includegraphics[width=0.8\columnwidth, right]{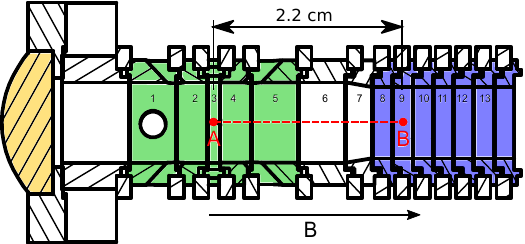}
	\caption{Cut section view of the electrodes used to transport the ion. Electrodes marked in green form the beryllium trap and electrodes marked in blue belong to the coupling trap. Electrode 3 is the four-split ring electrode of the beryllium trap, electrodes 2 and 4 the correction electrodes, and electrodes 1 and 5 the end caps. The lens collection light from fluorescence detection in marked in yellow. The magnetic field is indicated by the black arrow. The waveforms shown in figure \ref{fig:waveform} transport the ion from point A to point B.}
	\label{fig:transportpath}
\end{figure}

For harmonic confinement, the electric field at the ion position must be equal to 0 and the curvature is given by the desired trap oscillation frequency. The potential can also be constrained to a fixed value which makes equation \ref{eq:Aub} read \ref{eq:linsys} for $n$ electrodes

\begin{equation}
	\left( \begin{array}{rrrr}
		\phi_1(z)\vert_{z_0} & \phi_2(z)\vert_{z_0} & ... & \phi_n(z)\vert_{z_0} \\
		\frac{\partial \phi_1}{\partial z}(z)\vert_{z_0} & \frac{\partial \phi_2}{\partial z}(z)\vert_{z_0} & ... & \frac{\partial \phi_n}{\partial z}(z)\vert_{z_0} \\ 
		\frac{\partial^2 \phi_1}{\partial z^2}(z)\vert_{z_0} & \frac{\partial^2 \phi_2}{\partial z^2}(z)\vert_{z_0} & ... & \frac{\partial^2 \phi_n}{\partial z^2}(z)\vert_{z_0} \\
		
	\end{array}\right)
	\vec{u}\vert_{z_0}
	=\left(\begin{array}{c}
		V_0\\0\\\frac{m}{q} (2 \pi \nu_z)^2
	\end{array}\right)
	\label{eq:linsys}
\end{equation}
where $\phi_i$ is the potential at the ion position $z_0$ produced by the $i-$th electrode held at 1 V, $m$ and $q$ the ion's mass and charge, and $\nu_z$ the ion's axial trap frequency. $V_0$ is a potential constraint that can be chosen freely without loss of generality. We set it such that the potentials for ion transport match the potential we use for trapping an ion at position A (see fig. \ref{fig:transportpath}). For more than three electrodes, equation \ref{eq:linsys} is under-determined. To achieve the smallest possible voltages, the linear system of equations was solved using the Moore-Penrose inverse \cite{BenIsrael_2003} of $A$.

The potentials at the trap axis produced by an electrode at a voltage of 1 V while the other electrodes are held at 0 V were determined using finite element method (FEM) simulations with Comsol Multiphysics 4.2. For this purpose, the trap electrodes were approximated by hollow cylinders, i.e. the holes in the electrodes for laser access and ablation loading as well as the features for stacking the electrodes at the outside of the electrodes were omitted, and electrode no.~7 is implemented as a hollow truncated conical cylinder (see fig. \ref{fig:transportpath}) for the purpose of the simulations. Furthermore, holes for laser access and ablation loading as well as the gold plating of the electrodes were neglected.

\begin{figure}[b]
	\includegraphics[width=0.8\columnwidth, right]{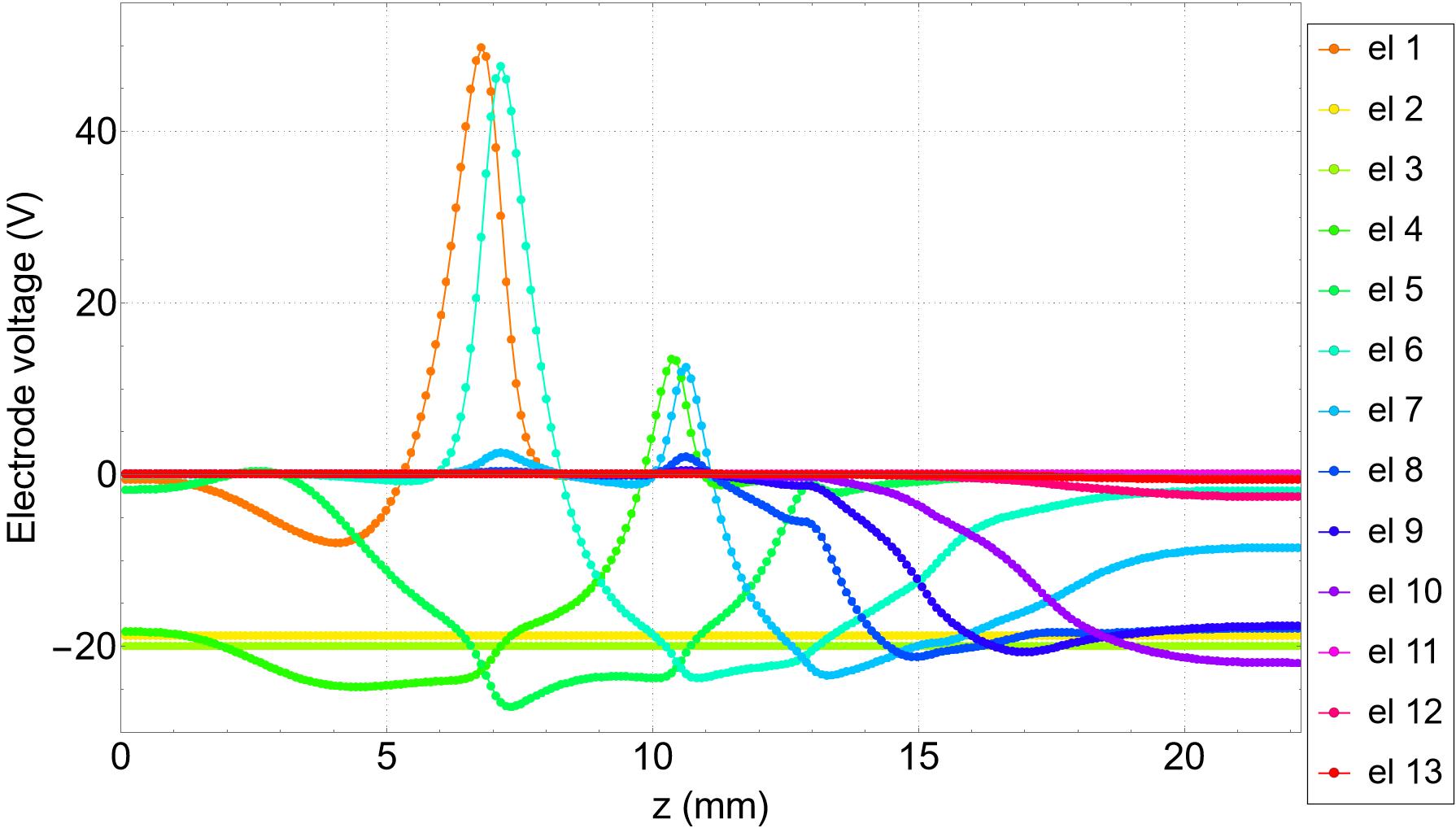}
	\caption{Voltage waveforms for transport of a beryllium ion at an axial trap frequency of 430 kHz over a distance of 2.2 cm along the trap axis $z$. Different colours denote different electrodes. For electrode reference see figure \ref{fig:transportpath}. The four-split ring electrode of the beryllium trap (labeled as el 3) as well as the correction electrode 1 (labeled as el 2) are constrained to -20 V and -18.8 V for the whole transport sequence to avoid perturbation of the ion. The lines serve as a guide to the eye.}
	\label{fig:waveform}
\end{figure}

In this work, waveforms for ion transport over a distance of 2.2\,cm have been calculated taking 13 electrodes into account. This allows to transport an ion from the beryllium trap to the coupling trap. Voltage vectors have been calculated using the method above for 196 positions along the transport path $z(t)$. The parametrization of $z(t)$ was chosen such that $\frac{\partial z(t)}{ \partial t} \sim \sin^2(t)$ between 0 and $\pi$ which provides a smooth start and stop of the transport process. Therefore, the positions were determined from eq.~\ref{eq:timedependence}
\begin{equation}
	z(t) = \frac{2(B-A)}{\pi}\Biggl[\frac{1}{2} \frac{\pi t}{T} - \frac{1}{4}\sin(2 \frac{\pi t}{T})\Biggr] + A
	\label{eq:timedependence}
\end{equation}
where $T$ is the duration for transporting an ion from the beryllium trap to the coupling trap, $A$ is the position on the $z$-axis where the ion is trapped prior to transport and $B$ the ion's target position. Eq. \ref{eq:timedependence} was evaluated for 196 equally spaced values of $t$ between $t=0$ and $t=T$. This parametrization provides a smooth start and stop of the transport process. Electrode 3 was held at a constant voltage during the whole transport sequence since it is split into four segments and possible issues from voltage differences between segments should be avoided. Electrode 2 was also held at a constant voltage during the transport sequence since it has a different low-pass filter compared to the rest of electrodes (see figure \ref{fig:electronics}). The resulting voltages for confining a single beryllium ion at an axial trap frequency of 430 kHz at different positions along the trap axis are displayed in figure \ref{fig:waveform}.

\section{Experimental setup}
\label{sec:setup}

Figure \ref{fig:transportpath} shows a cut section view of the beryllium trap (electrodes 1 to 5), some electrodes of the coupling trap (electrodes 8 to 13) and the transport electrodes (electrodes 6 and 7). The Penning trap stack consists of gold plated copper electrodes that are electrically isolated from each other by sapphire rings. The hole in electrode 1 (see fig. \ref{fig:transportpath}) holds a beryllium target to enable laser-ablation loading of beryllium ions. Furthermore, electrodes 1 and 5 contain holes for laser beams. For a full description of the current trap stack see \cite{meiners_characterisation_2021}. The waveforms shown in figure \ref{fig:waveform} are suitable for transporting a single beryllium ion from the center of the beryllium trap (A) to the coupling trap (B) and back, which corresponds to a total distance of 4.4\,cm.

\begin{figure}[t]%
	\centering
	\includegraphics[width=0.8\columnwidth, right]{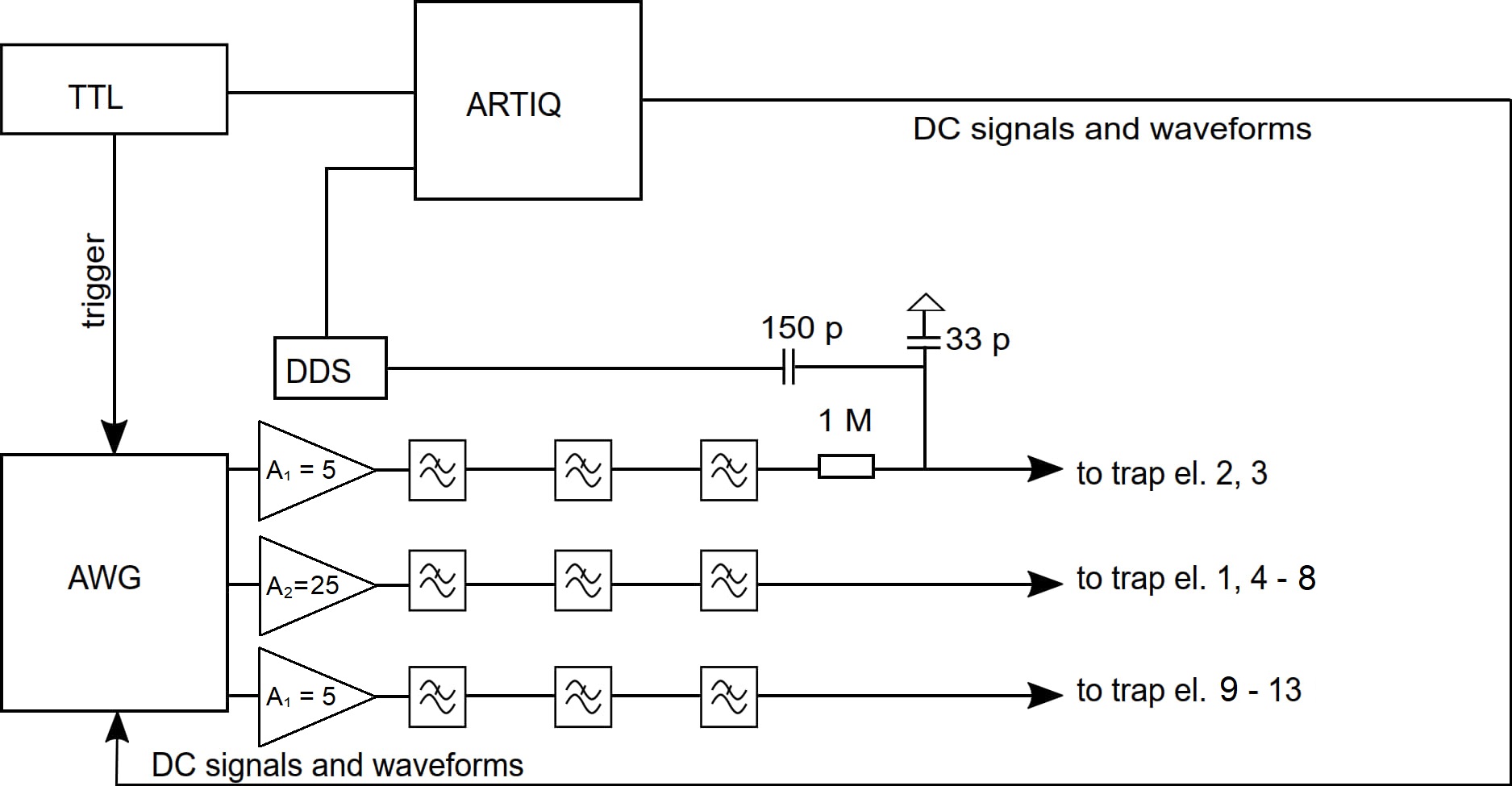}
	\caption{Schematic of the control electronics and trap wiring. The electrode voltages are generated by an AWG. The signals for electrodes 2, 3 and 9 to 13 are amplified by a self-built amplifier with a gain factor of 5. The signals for electrodes 1 and 4 to 8 are amplified by another amplifier with a gain factor of 25. All DC signals are filtered by a three staged low-pass filter with capacities of 4.7 nF and resistances of 5 k$\Omega$. Electrodes 2 and 3 are additionally connected to a direct digital synthesis (DDS) module that generates RF signals for motional excitation of the ion. Both the AWG and DDS module are controlled by the experiment control system ARTIQ. For details see text.
	}
	\label{fig:electronics}
\end{figure}

The control electronics and the trap wiring are depicted in figure \ref{fig:electronics}. The voltages for trapping and transporting the ion are generated by a programmable arbitrary waveform generator (AWG) developed at NIST \cite{bowler_2013}. It produces DC voltages in a range of $\pm10$\,V with 16 bit resolution with an update rate of 50 MHz. The signals for electrodes 2, 3 and 9 to 13 are amplified to $\pm30$\,V by a voltage amplifier. The signals for electrodes 1 and 4 to 8 are amplified to $\pm250$\,V by a commercial low-noise high-voltage amplifier (PiezoDrive TD250). All the signals pass three cascaded low-pass filters with a resulting cutoff frequency of 750\,Hz for electrodes 2 and 3 and 1.3\,kHz for the remaining electrodes before they reach the electrodes.

The transport voltages were rounded to the level of $1 \cdot 10^{-4}$ V due to the resolution of the AWG.  
The voltages for each electrode were subsequently interpolated by a third order b-spline. The interpolation coefficients as well as the voltages were handed over to the AWG where voltage waveforms with a resolution of 20 ns are generated \cite{jordens_pdq_2018}. 

The AWG is controlled by the control system ARTIQ (Advanced Real-Time Infrastructure for Quantum physics) from m-labs \cite{m-labs_notitle_2022}. It is used for programming the AWG as well as for starting the ion transport by sending a trigger signal to the AWG. Since ARTIQ executes commands in real time and also controls switching of the cooling and detection laser, laser and ion interaction can be controlled in real time.

\section{Experimental verification}
\label{Sect:Experimental verification}

As a first demonstration, we verify the transport of a single $^9$Be$^+$ ion confined in the cryogenic Penning traps system. More details about the experimental setup can be found in \cite{Niemann_2019}. To summarize, once ions have been loaded into the beryllium trap using an ablation laser pulse at 532~nm impinging on a beryllium target embedded into electrode 1 (see figure \ref{fig:transportpath}), a single ion is obtained by splitting the ion cloud several times. After this process, eventually a single $^9$Be$^+$ ion is maintained in the beryllium trap. By using a laser beam 10~MHz red-detuned from the cooling transition ($^2S_{1/2}\ket{m_I=3/2,m_J=1/2}$  $\leftrightarrow$ $^2P_{3/2}\ket{m_I=3/2,m_J=3/2}$) of the $^9$Be$^+$ ion and an offset from the trap center in the vertical direction, an initial axial mode temperature of $ T_{z} \approx$~2~mK is obtained for the single ion by Doppler cooling. For an ion cloud, a temperature of 1.77~(10)~mK has been measured in our experiment \cite{Mielke_2021}. By using the same technique as in ref. \cite{Mielke_2021}, a similar value is obtained for the single ion. This value is close to the expected Doppler cooling limit. For a discussion of our experimental boundary conditions for cooling, see \cite{Mielke_2021}. Besides the Doppler laser, a repumper laser on resonance with the repumping transition ($^2S_{1/2}\ket{m_I=3/2,m_J=-1/2}$ $\leftrightarrow$ $^2P_{3/2}\ket{m_I=3/2,m_J=1/2}$) is used to avoid undesirable quantum jumps \cite{Bergquist_1986}.

For the measurements shown in this section, the single $^9$Be$^+$ ion is stored by using a trap bias of -20~V on the ring electrode and -18.8~V and -18.9~V  on electrodes 2 and 4 (see figure~\ref{fig:transportpath}) which are the correction electrodes, respectively. 
These values yield an axial frequency of $\omega_{z}/2 \pi \approx$~430~kHz, and magnetron and reduced cyclotron frequencies of about 10 kHz and 8,490 kHz, respectively \cite{mielke_thermometry_2021} in the beryllium trap. For cooling, 300~$\mu$W of power for the Doppler laser is used with a beam waist of around 150~$\mu$m. Only 30~$\mu$W are used for the repumper laser with a similar beam waist. In addition, an offset position for the Doppler laser of around 100~$\mu$m with respect to the trap center is used in order to cool both radial degrees of freedom. The fluorescence signal from the cooled ion is detected by a photomultiplier tube, whose output signal is connected to ARTIQ in order to register the photons count. 

\begin{figure}[t]
	\includegraphics[width=0.8\columnwidth, right]{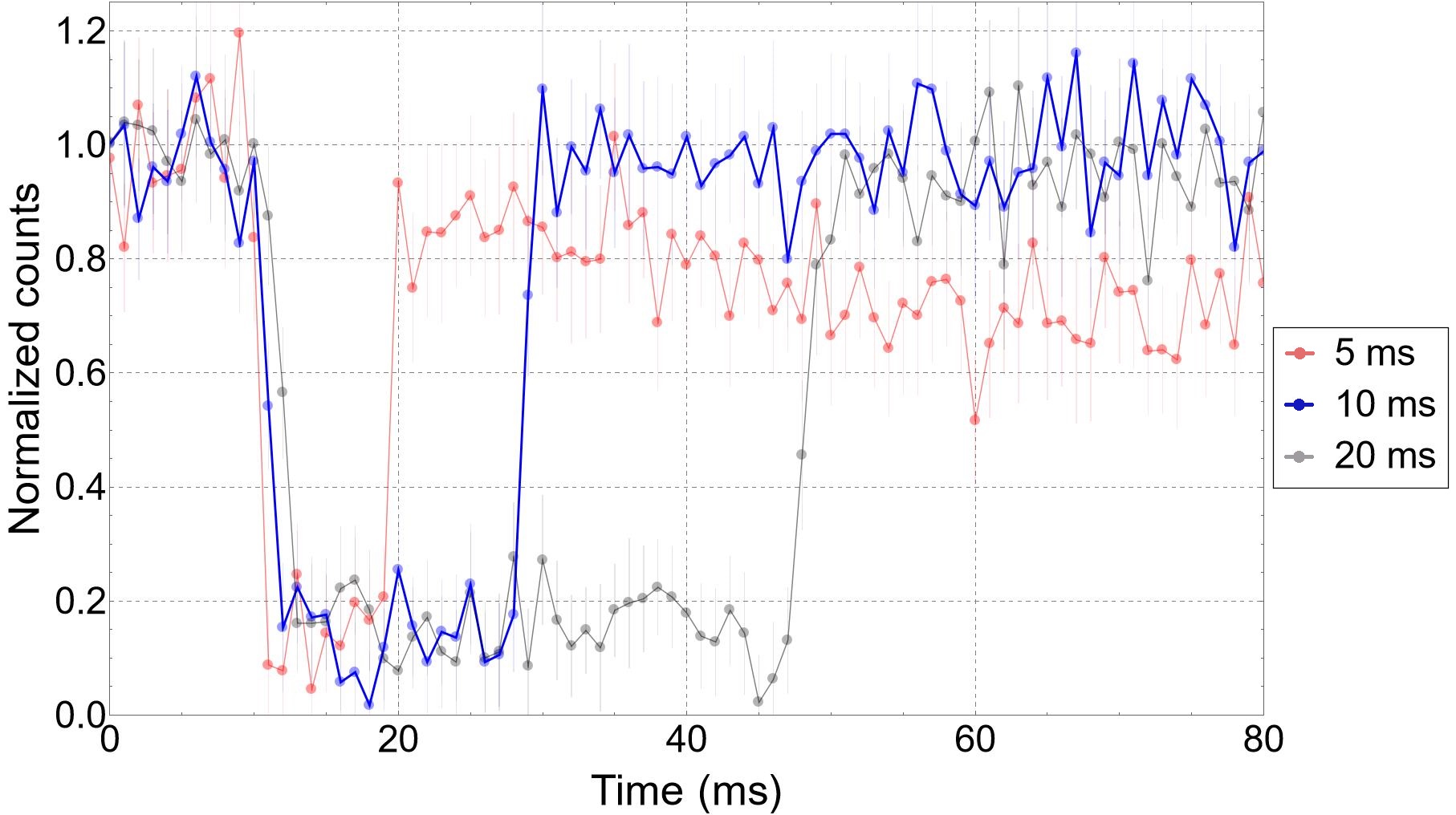}
	\caption{Observed fluorescence counts as a function of time during ion transport for different shuttling times. The experimental data for a shuttling time of 5~ms, 10~ms and 20~ms are represented by solid red, blue and black circles, respectively. Error bars are based on counting statistics. 50 repetitions were used for each transport sequence. Solid lines are shown as a guide to the eye. An exposure time of 800~$\mu$s was used for each data point with a delay time between points of 200~$\mu$s. The counts are normalized to the maximum observed for the 10~ms shuttle-time. Further details in the text.}
	\label{fig:data}
\end{figure}

\begin{figure}[t]
	\centering
	\includegraphics[width=0.8\columnwidth, right]{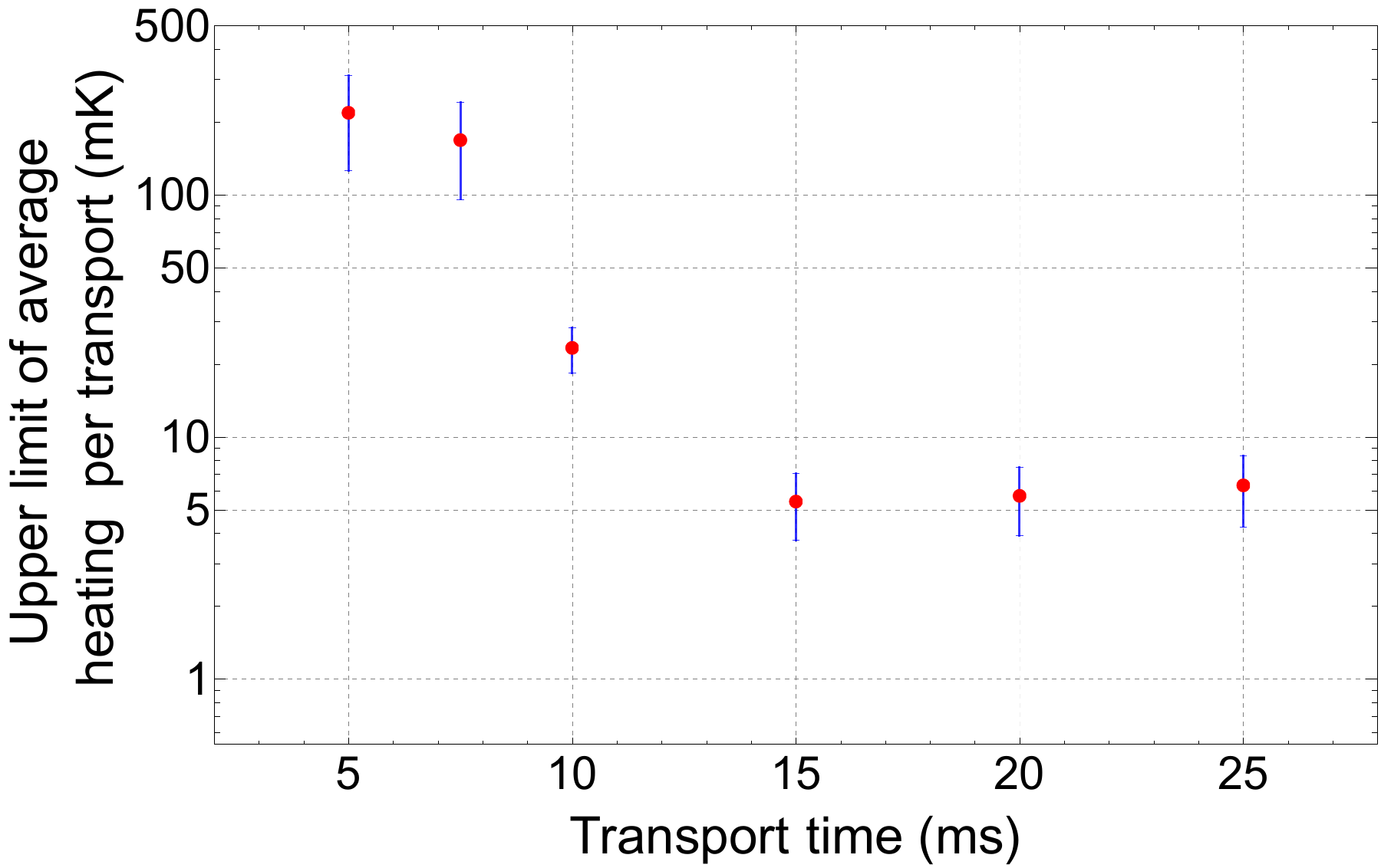}
	\caption{Upper limit of the average heating per transport for several transport times. The ion heating after several transports were measured by recooling for different shuttle times. For shuttle times of 10~ms, 15~ms, 20~ms and 25~ms, 100 transports repetitions were performed before measuring the ion heating. For shuttle times of 5~ms and 7.5~ms, 5 and 20 transports repetitions were performed. The mean ion energy after transport was obtained by fitting the normalized average moving counts over time for single measurements. Error bars are based on 50 measurements repetitions, except for a shuttle time of 5~ms where only 10 measurements repetitions could be realized. Further details in the text.}
	\label{fig:data2}
\end{figure}

In order to test the transport waveforms shown in figure~\ref{fig:waveform}, a single ion was transported from the beryllium trap (point A in figure~\ref{fig:transportpath}) to the coupling trap (point B in figure~\ref{fig:transportpath}), and transported back to the beryllium trap. Figure~\ref{fig:data} shows the normalized counts observed in the beryllium trap, where the ion was first cooled during 10~ms, then transported back and forth, and finally cooled again in the beryllium trap. Different shuttling times for moving the ion from one trap to another have been considered. The lasers were turned on during the entire process. Therefore, the drop in fluorescence observed around 10 ms in figure~\ref{fig:data} demonstrates the absence of the ion in the beryllium trap as a result of the transport. Since the transport waveforms are applied twice (once forward and once reversed) due to the round trip, the low fluorescence level is observed approximately twice as long as that of the shuttle time. Although the measurements shown in figure~\ref{fig:data} demonstrate transport on the expected time scale, they do not demonstrate transport of the ion to zone B (see figure~\ref{fig:transportpath}). To ensure that the latter has occurred, for each transport the waveform to move the ion from zone A to zone B is first applied. Then the potential in zone A is changed to the trapping potential. Then the potential in zone A is modified to the initial transport potential and finally the transport waveform to move the ion from zone B to zone A is applied. Since it is only possible to trap the ion in zone A and zone B, and the ion is only detected after to apply the waveform to move the ion from zone B to zone A, for each transport waveform, we can ensure transport of the ion to zone B.

Before transport, Doppler cooling prepared the ion in a thermal state (\cite{mavadia_optical_2014}) with a mean phonon number in the axial mode given by $k_{B}T_{z}/(\hbar\omega_{z}) \approx$~100, where $k_B$ is the Boltzmann constant and $\hbar$ is the reduced Planck constant. After transport, the recooling process can be examined in order to estimate the mean ion energy. By using this technique, the thermally averaged scattering rate over time can be used in order to calculate the mean ion energy before recooling~\cite{wesenberg_fluorescence_2007}. In this model, we consider a Maxwell-Boltzmann distribution of the motional energies $\varepsilon$ given by $P(\varepsilon) = \exp[-\varepsilon/\bar{\varepsilon}]/\bar{\varepsilon}$ for each cooling period, where $\bar{\varepsilon}$ is the mean energy. From this distribution, a thermally averaged scattering rate \cite{wesenberg_fluorescence_2007}  
\begin{equation}
	\label{rate}
	\left< \frac{dN}{d\tau} \right>_{\bar{\varepsilon}} = \left. \int_{0}^{\infty} \frac{e^{-\varepsilon^{\prime}/\bar{\varepsilon}}}{\bar{\varepsilon}} \frac{dN}{d\tau}\right|_{\varepsilon = \Xi(\varepsilon^{\prime}, \tau)} d\varepsilon^{\prime}
\end{equation} 
is obtained, where $dN/d\tau$ is the scattering rate for a given motional energy $\varepsilon$ and $\Xi$ is the propagator of $\varepsilon$, i.e. the energy at time $\tau$ of an ion with initial energy $\varepsilon_0$. 

Non-adiabaticity in transport can in principle produce coherent effects on the motion and thus non-thermal states. In our case, the transport time is still 4 orders of magnitude longer than the trap oscillation period, so that some fluctuations superimposed on any coherent effects could be expected. Note that this is a common question, which in principle also affects heating rate measurements on stationary ions, where technical sources of resonant fields can lead to coherent effects. The Doppler recooling technique with its sensitivity to Doppler shifts should still give a decent measure of energy.

In order to calculate the thermally averaged scattering rate over time given by Eq.~\ref{rate}, we consider our trapped $^9$Be$^+$ ion with a cooling transition at 313~nm, a natural linewidth of $\Gamma = 2 \pi \times$ 19.6~MHz, a laser detuning from resonance of $-2 \pi \times$ 10~MHz, a saturation parameter of $s = 0.05$ and a wavevector $k_z/k = 0.5$. Since our laser also has a radial component, this model can be used in order to determine an upper limit of the axial heating per transport. Longer recooling times due to radial effects affect the measurements by overestimating energy gains in the axial mode. Taking into account the above values, an upper limit of the mean energy of the ion is determined from the scattering rate model. A further discussion of the recooling process and the smallest energy which is observable with binning time can be found in Appendix~\ref{Sect:Appendix: A}. From the data discussed above, an upper limit for the ion temperature before recooling of $\approx$~1~K can be estimated, but no recooling process is observed. This is due to the chosen binning time and because of the transport through the spatially inhomogeneous laser beam.

More accurate temperature information can be extracted by turning the laser beams off prior to transport and turning them back on only after the round-trip transport has been completed, and by examining specifically the fluorescence during the recooling process only. To accomplish this, the ion was first cooled during 100~$\mu$s, then the lasers were turned off and the ion was transported back and forth a total of 100 times to amplify any potential energy gain from transport. After that, the lasers were turned on and the fluorescence signal was recorded. Each recooling signal was fitted using Eq.~\ref{rate} and the fitted temperature was divided by the number of transports. 
This process is repeated for each shuttle time in order to improve the statistics. It is important to note that when no heating is observed, then a final temperature of 1~K is taken (see above and appendix). 

Figure~\ref{fig:data2} shows the upper limit of the average heating per transport for several shuttle times. For shuttle times of 15~ms, 20~ms and 25~ms a similar heating is observed, with values lower than 10~mK per transport.
For shuttle times of 5~ms, 7.5~ms and 10~ms, pronounced heating can be observed. This is the expected result when the time constants of the RC low-pass filters of the trap electrodes are considered. The low-pass filters introduce a significant delay between the signals applied to the low-pass filters and the voltages appearing on the electrodes, thus introducing significant distortions of the actual temporal evolution of the trapping fields compared to the intended behavior. This effect is even more important for the shorter shuttle times of 5~ms and 7.5~ms, where the number of transport repetitions was reduced from 100 to 5 and 20, respectively. In these cases, for higher repetition counts, it could be observed that occasionally the ion would be in a higher magnetron orbit and even appear to be “missing”, because at a sufficiently high magnetron orbit our experimental system does not allow us to Doppler cool the ion again \cite{Niemann_2019}. These distortions may also be responsible that the fluorescence signal does not return to the original level within the observed time frame for the 5~ms shuttle time in figure~\ref{fig:data}, potentially also connected with an effect on the radial motion. Note that the measurement is carried out after one or several round-trips, which would in principle allow for the motion to be coherently excited on the way in and then de-excited on the way back. If this were the case, it could be detected e.~g.~ through the introduction of a variable waiting time between the trip in and back~\cite{todaro_improved_2020}. We have not carried out such measurements, but it would be plausible that such an effect, if it were pronounced, would show up for at least one of the 15, 20 and 25\,ms transport times and not conspire to cancel out perfectly each time. 

\section{Summary and Outlook}

Successful transport of a single laser cooled $^9$Be$^+$ ion over 2.2~cm in a Penning trap has been demonstrated on timescales in the ms regime. This is fast compared to ion transport previously demonstrated in Penning traps where transport durations are on the order of seconds (see e.g. \cite{heisse2017high} and \cite{smorra_350-fold_2018}). For transport times of 15, 20 and 25 ms, energy gains below 10~mK were observed using the Doppler recooling method. Shorter transport times can introduce significant amounts of heating, expected from the distortion of the control signals through the low-pass filters. We show the first study of the effect of fast adiabatic transport on the axial degree of freedom in a Penning trap. 

Present efforts to implement sympathetic laser cooling in Penning trap systems will lead to significantly reduced cooling and readout times. The development of fast adiabatic transport procedures is a key step to realize the full potential of sympathetic laser cooling such that the present transport times of e.g.~minutes
do not limit the measurement time and such that energy gains from transport do not spoil the effect of the sympathetic cooling. 

Further insights into transport-induced heating can be obtained by measuring the energy gain of the ion by first applying axial ground state cooling, followed by transport and sideband spectroscopy~\cite{Mielke_2021}. Future improvements to the experimental setup in terms of ion detection and cooling would allow us to further increase the transport speeds by pre-distorting the transport waveforms as described in e.g. \cite{todaro_improved_2020} to account for the transfer functions of the low-pass filters.

These results also represent a significant step towards the implementation of quantum logic spectroscopy in Penning traps. Here, the axial degree of freedom is used for the required energy and information exchange between the particle of interest (spectroscopy particle) and the cooling ion. Transport is required in such a scenario because typically the requirements for laser cooling, precision spectroscopy and energy exchange between cooling ion and spectroscopy particle cannot be fulfilled in the same trapping site, necessitating the implementation of a multi-well trap stack interconnected with fast adiabatic transport as demonstrated here. For a discussion of a possible implementation in the \ppbar system, consider \cite{Cornejo_2021}. 

\section*{Acknowledgments}
We acknowledge financial support from DFG through SFB/CRC  1227 ‘DQ-mat’, project B06 and through the cluster of excellence QuantumFrontiers, from the RIKEN Chief Scientist Program, RIKEN Pioneering Project Funding, the Max Planck-RIKEN-PTB Center for Time, Constants, and Fundamental Symmetries, and the European Research Council (ERC) under FP7 (grant Agreement No. 337154).

\section*{Conflict of interest}
The authors declare no competing interests

\section*{Data availability statement}
The data that support the findings of this study are available upon reasonable request from the authors.

\section*{Author contribution}
TM and JMC have calculated the waveforms for ion transport. TM, JAC and MN have experimentally implemented ion transport in our setup. JMC and JAC have performed the measurements for single ion transport. JM has designed, implemented and characterised the laser systems. SU, MN, TM, JM and JMC have worked on the construction of the Penning trap stack. SU and CO have guided the project. CO initiated and supervised the project. All authors have discussed the findings of the manuscript.

\begin{appendices}

\begin{figure}[t]
	\includegraphics[width=0.7\columnwidth, right]{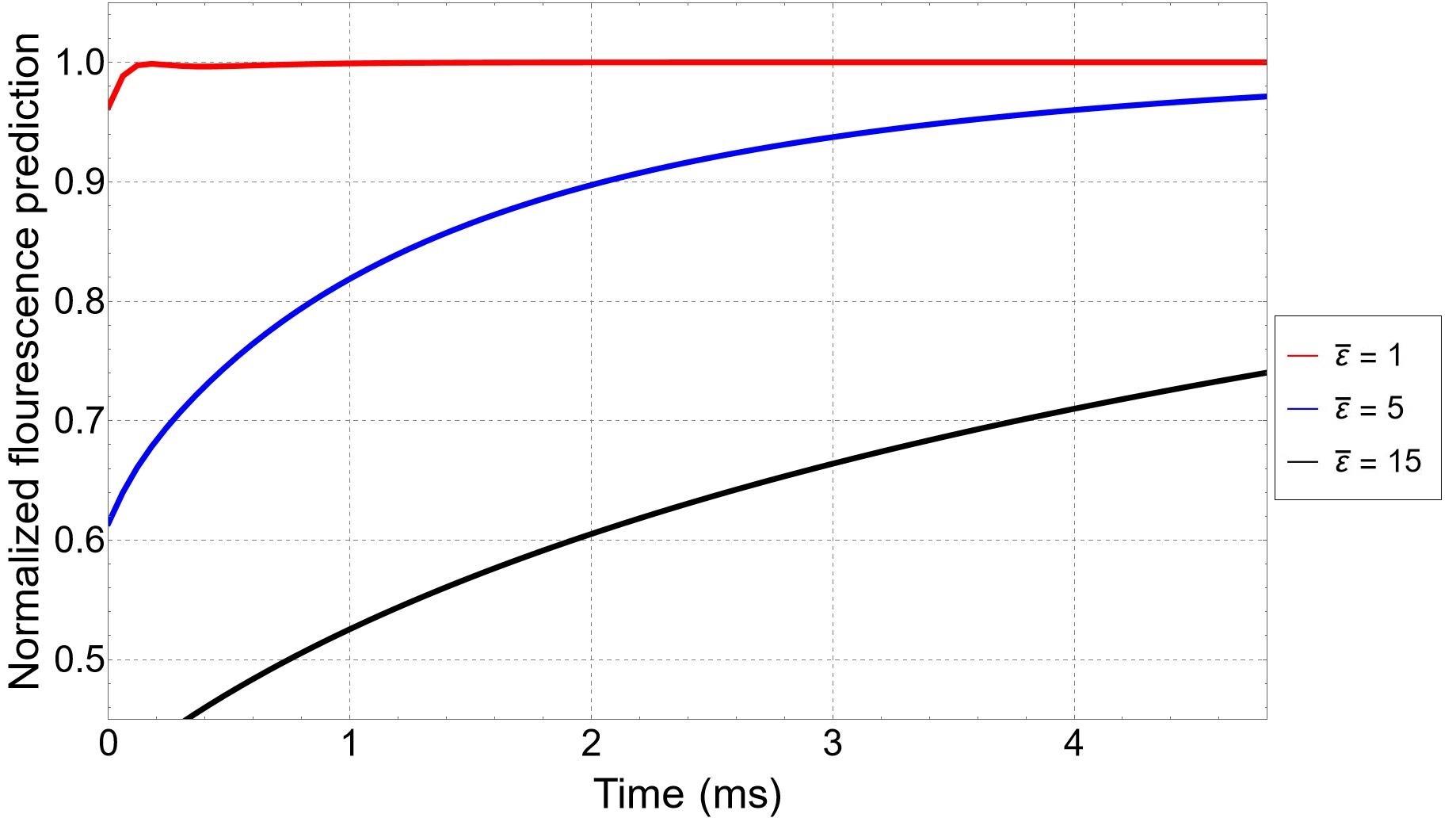}
    \caption{Normalized fluorescence predicted over time for several mean ion energies. Red, blue and black lines represent the thermally averaged scattering rate over time for mean energy values of 1, 5 and 15, respectively.}
	\label{fig:mean_energy}
\end{figure}

\begin{figure}[b]
	
	\includegraphics[width=0.7\columnwidth, right]{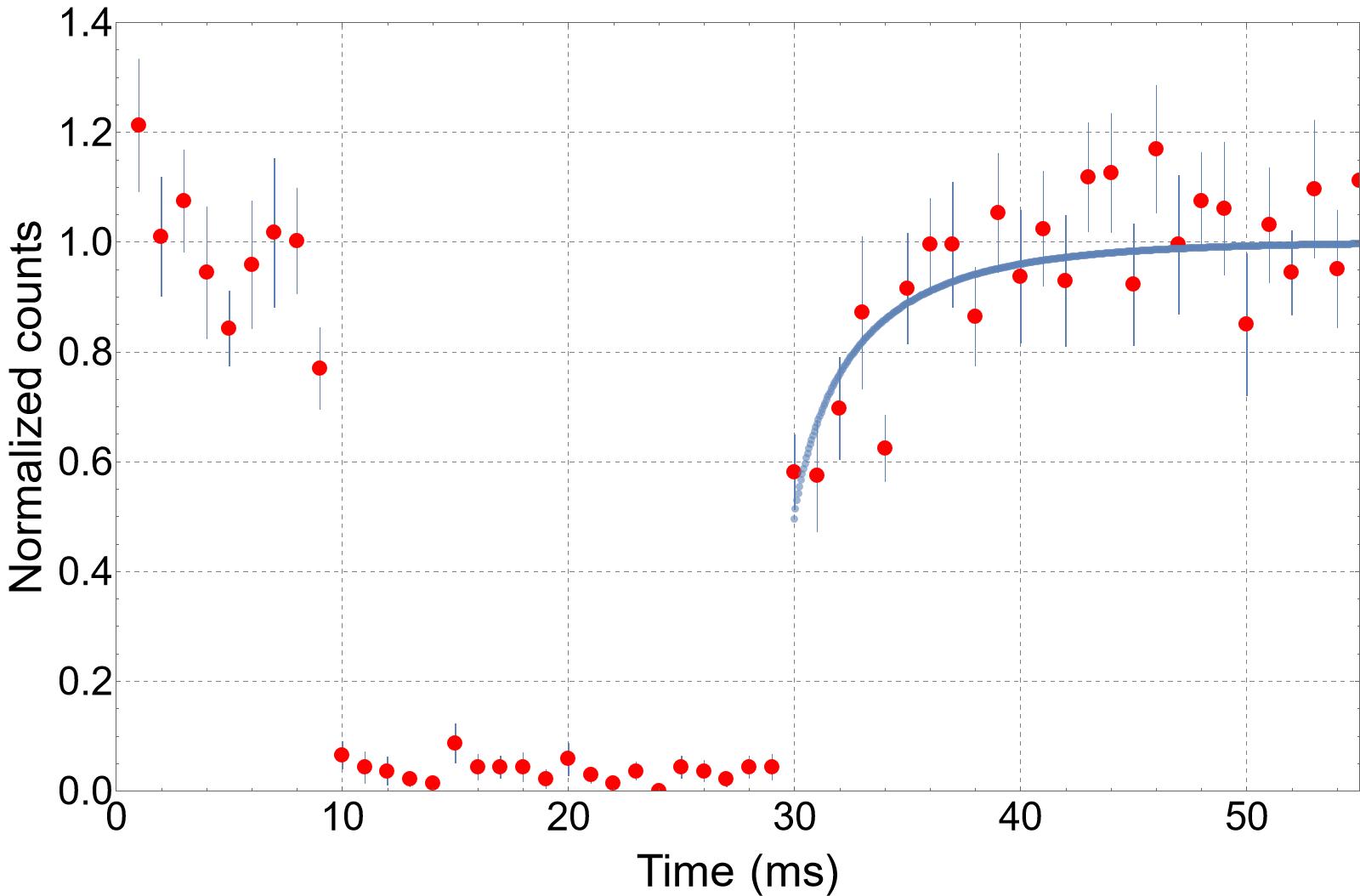}
	\caption{Normalized counts over time for a heated ion. The data is represented by red dots and error bars are based on counting statistics. 20 repetitions were performed. Blue line is a thermally averaged scattering rate fit to the data. The fit gives a temperature value of 1.5~K after heating.}
	\label{fig:exc}
\end{figure}

\section*{Appendix}
\label{Sect:Appendix: A}

In order to determine the minimum ion energy that is observable with our experimental parameters by using the thermally averaged scattering rate over time during recooling, the predicted florescence was calculated for several mean energy values. This is shown in Figure~\ref{fig:mean_energy} for mean ion energies of 1, 5 and 15. Main energies are given in recoil units $r = (\hbar k_z)^2 / 2 m)/E_0 \approx 0.0055$, where $m$ is the ion mass and $E_0 = \hbar \Gamma \sqrt{1+s}/2$. The relation between temperature and main energy is given by $T = 2 E_0 \bar{\varepsilon} / r k_B $. In this model, time units are given by $t_0 = 2(1+s)/s \Gamma \approx$~0.179~$\mu s$.

Taking into account the recooling time as a function of the initial energy (see figure~\ref{fig:mean_energy}) as well as the binning time of 1\,ms chosen for photon counting during recooling, we can consider a mean energy value of around 5 as the minimum energy that can be observed. This energy is equivalent to an ion temperature of about $\approx$~1~K.

In order to check that the implementation of the method is capable of detecting axial energy gains, the single ion was axially excited with a purely axial radiofrequency excitation close to the axial frequency. Note that this is expected to produce a coherent state of motion. Figure 7 shows the counts over time, where the ion is laser cooled without excitation, then the laser is turned off and the excitation is turned on for 20~ms, and finally, the excitation is turned off and the Doppler laser is turned on again. In this case, the recooling process can be observed and the predicted averaged scattering rate can be fitted to the data, corresponding to an ion mean energy before the recooling of 8.8, or a temperature of 1.5~K.

\end{appendices}

\bibliographystyle{jphysicsB}
\bibliography{transport_paper}

\end{document}